\input phyzzx

\font\medium=cmbx10 scaled \magstep1
\font\large=cmbx10 scaled \magstep2
\vsize 23.5 cm
\hsize 16.5 cm
\hoffset 0.1 cm
\voffset 0 cm
\line{\hfill\hbox{NCKU-HEP/96-05}}
\vskip 3 cm
\centerline{\large Split Dimensional Regularization for the Temporal Gauge}
\vskip 1cm
\centerline{Yaw-Hwang Chen, Ron-Jou Hsieh}
\vskip 0.2cm
\centerline{\it and}
\vskip 0.2cm
\centerline{Chilong Lin}
\vskip 0.4 cm
\address{Department of Physics, National Cheng Kung University}
\address{Tainan, Taiwan 701, Republic of China}

\abstract
\noindent
A split dimensional regularization, which was introduced for the 
Coulomb gauge by Leibbrandt and Williams, is used to regularize the 
spurious singularities of Yang-Mills theory in the temporal gauge.  
Typical one-loop split dimensionally regularized temporal gauge integrals, 
and hence the renormalization structure of the theory are shown to be the 
same as those calculated with some nonprincipal-value prescriptions.  
\filbreak
\vfill \eject

In the studies of gauge theories, we are generally required to choose 
a gauge for quantization.  Among the feasible gauge conditions, 
noncovariant gauges [1,2], which can be classified by constant 
four-vectors, have been most discussed from the technical point of view.  
Among the noncovariant gauges, the temporal gauge is known to be the 
most complicated and cumbersome.  Nonprincipal value prescriptions [3,4] 
have been applied to study the renormalization of Yang-Mills 
theory quantized in the temporal gauge.  
These prescriptions provide useful 
calculational procedures for the dimensionally regularized temporal-gauge 
integrals.  Recently, a regularization known as the split dimensional 
regularization was proposed by Leibbrandt and Williams [5] for 
Yang-Mills theory in the Coulomb gauge.  In this brief report, we shall apply 
the split dimensional regularization to study the renormalization of 
Yang-Mills theory in the temporal gauge. 

The temporal gauge is defined by the condition $n_\mu A_{\mu}^a 
= A_{0}^a = 0$, where $A_{\mu}^a$ is the gauge potential and $n_\mu 
= (1,0,0,0)$ a constant temporal four-vector.  In this gauge, the propagator 
has a spurious double pole at $q_{_0} = 0$ and reads $(i,j = 1,2,3)$
$$
G_{ij}^{ab}(q) = {{i\delta^{ab}}\over{q^2 + i\epsilon}}\left\{
\delta_{ij} + {{q_iq_j}\over{q_{_0}^2}}\right\},\,\,
G_{0i}^{ab}(q) = G_{i0}^{ab}(q) = G_{00}^{ab}(q) = 0,\eqno(1)
$$
or, in covariant form,
$$
G_{\mu\nu}^{ab}(q) = {{i\delta^{ab}}\over{q^2 + i\epsilon}}\left\{
- \delta_{\mu\nu} + {{(q_\mu n_\nu + q_\nu n_\mu)}
\over{q\cdot n}}
- {{q_\mu q_\nu}
\over{(q\cdot n)^2}}\right\}\,\,,\,\epsilon > 0\,,\eqno(2)
$$
where we use a $(+,-,-,-)$ metric.  We note that propagator (2)
has a simple pole and a double pole at $q_{_0} = 0$.
 
Our purpose of this letter is to outline the calculational procedure for the
one-loop regularized temporal-gauge integrals and to observe the gauge 
divergence problem of temporal-gauge theories.  At the one-loop level, we 
shall need a regularization for regularizing the gauge divergences of
the integrals.  For the usual ultraviolet divergences that we are interested 
in, we employ split dimensional regularization with complex space-time 
dimensionality $2(\omega + \sigma) \equiv D$, with $2\omega$ space dimensions
and $2\sigma$ time dimensions.

We first consider the following integral with the spurious simple pole
in Euclidean space:
$$
I = \int {{d^Dq}\over{(p - q)^2q\cdot n}}\,.
\eqno(3)
$$
Using Feynman's parameterization and exponentiation formulae, we get
$$
\eqalignno{
I &= i\int_0^1 dx\int {{d^Dq\,\,\,\,
q_0}\over{\left[x\left((p - q)^2 + i\epsilon
\right) + (1 - x)q_0^2
\right]^2}}\cr
&= i\int_0^1 dx\int {d^Dq}\,
{{q_0 }\over{\left[q_0^2 +
x\vec{q}^2 + xp^2 - 2xp_0q_0 - 2x\vec{p}\cdot\vec{q}
\right]^2}}\cr
&= i\int_0^1 dx \int_0^{\infty} d{\alpha} \alpha e^{-\alpha x p^2}\,
\int d^{2\omega}\vec{q} e^{-\alpha(x\vec{q}^2 - 2x\vec{p}\cdot\vec{q})}
\int d^{2\sigma} q_0 q_0 e^{-\alpha(q_0^2 - 2xp_0 q_0)}\,.&(4)\cr}
$$
Because
$$
\int d^{2\omega}\vec{q} e^{-\alpha x(\vec{q}^2 - 2\vec{p}\cdot\vec{q})}
 = \pi^{\omega} (\alpha x)^{-\omega} e^{\alpha x \vec{p}^2}\,,
\eqno(5)
$$
$$
\int d^{2\sigma}q_0 q_0 e^{-\alpha (q_0^2 - 2xp_0q_0)}
= \left[{{\pi^{\sigma + {1\over 2}} (2\sigma - 1)!!} \over {\Gamma(\sigma)
2^{\sigma} \alpha^{\sigma + {1\over 2}}}}
+ xp_0 \pi^{\sigma} \alpha^{-\sigma}\right] e^{\alpha x^2 p_0^2}\,,
\eqno(6)
$$
we obtain
$$
\eqalignno{
I &= i\pi^{\omega + \sigma} p_0\int_0^1 dx x^{1-\omega}
\int_0^\infty d\alpha \alpha^{1-\omega - \sigma} e^{-\alpha x(1-x)p_0^2}\cr
&= i\pi^{\omega + \sigma}p_0 \Gamma(2-\omega - \sigma) \int_0^1 dx
x^{1 - \omega}\cr
&= {{2p\cdot n}\over{n^2}}{\overline I} \,\,,\,\,
{\overline I} \equiv i\pi^2\Gamma(2 - \omega -\sigma)\,\,,\,
\omega \rightarrow {3\over 2},\, \sigma \rightarrow {1\over2} \,.&(7)\cr}
$$
This result is the same as that for the corresponding temporal gauge
integral calculated with some nonprincipal-value prescriptions.

Next we turn to an integral with the double pole
in Euclidean space:
$$
J = \int {{d^Dq}\over{(p - q)^2(q\cdot n})^2}\,.
\eqno(8)
$$
Following the same procedure for the previous integral, we get
$$
\eqalignno{
J &= i\int_0^1 dx\int {{d^Dq\,\,\,\,\,\,
}\over{\left[x\left((p - q)^2 + i\epsilon
\right) + (1 - x)q_0^2
\right]^2}}\cr
&= i\int_0^1 dx\int {d^Dq}\,{1\over{\left[q_0^2 +
x\vec{q}^2 + xp^2 - 2xp_0q_0 - 2x\vec{p}\cdot\vec{q}
\right]^2}}\cr
&= i\int_0^1 dx \int_0^{\infty} d{\alpha} \alpha e^{-\alpha x p^2}\,
\int d^{2\omega}\vec{q} e^{-\alpha(x\vec{q}^2 - 2x\vec{p}\cdot\vec{q})}
\int d^{2\sigma} q_0 e^{-\alpha(q_0^2 - 2xp_0 q_0)}
\,.&(9)\cr}
$$
Performing the q-integral, we obtain
$$
\eqalignno{
J &= i\pi^{\omega + \sigma} \int_0^1 dx x^{-\omega}
\int_0^\infty d\alpha \alpha^{1-\omega -\sigma} e^{-\alpha x(1-x)p_0^2}\cr
&= i\pi^{\omega + \sigma} \Gamma(2-\omega - \sigma) \int_0^1 dx
x^{-\omega}\cr
&= {-2\over{n^2}}{\overline I}\,,
\omega \rightarrow {3\over 2},\,\sigma\rightarrow{1\over2}\,.&(10)\cr}
$$
which is the same as the corresponding temporal integral calculated with 
some prescriptions [3,4].  Other temporal-gauge integrals needed
for the one-loop gluon self-energy can be easily calculated (cf. ref. [4]).

We next briefly mention the renormalization of the temporal gauge theory 
in split dimensional regularization.  We consider the one-loop gluon 
self-energy.  Let the time-translation invariant gauge propagator [4] be:
$$
G_{\mu\nu}^{ab}(q) = {{i\delta^{ab}}\over{q^2 + i\epsilon}}\left(
-\delta_{\mu\nu} + a_\mu(q)q_\nu- a_\nu(-q)q_\mu\right)\,,\eqno(11)
$$
where $a_\mu(q)$ is an arbitrary function related to the gauge choice.
Let the ghost-gluon-ghost vertex be:
$$
{{\Gamma_\mu}^{\!\!\!abc}}(q) =
- gf^{abc}\left(\left[(a\cdot q) - 1\right]q_\mu - q^2
a_\mu(q)\right)\,,\eqno(12)
$$
and let the ghost propagator be:
$$
G(q) = {{-i}\over{{q^2 + i\epsilon}}}\,,\,\,\epsilon > 0\,,\eqno(13)
$$
with $g$ being the coupling constant, $q_\mu$ being the outgoing ghost's
momentum.
We get for the temporal gauge
$$
a_\mu(q) = {{n_\mu}\over{q\cdot n}} - {{q_\mu}\over{2(q\cdot
n)^2}}\,,\eqno(14)
$$
$$
{{\Gamma_\mu}^{\!\!\!abc}}(q) =
gf^{abc}{{q^2n_\mu}\over{q\cdot n}}\,.\eqno(15)
$$
obtained by the Faddeev-Popov gauge-fixing procedure.  Using the procedure,
we have a ghost-gluon-ghost vertex that is proportional to $n_\mu$ and a
ghost propagator that is inversely proportional to $q\!\cdot\!n$.  
It is easy to show that in split dimensional regularization 
the one-loop ghost diagram vanishes.
Therefore, the calculation of the one-loop gluon self-energy requires 
considering one diagram with an internal gluon loop.

The calculation of the divergent part of the one-loop gluon self-energy
has been carried out and yields
$$
i{\Pi_{\mu\nu}}^{\!\!\!\!\!\!ab}(p) =
{{11g^2C_A}\over{3(2\pi)^{2(\omega +\sigma)}}}\delta^{ab}(p^2\delta_{\mu\nu}
- p_\mu p_\nu){\overline I}\,,\eqno(16)
$$
where $C_A = N$ for $SU(N)$ gauge group.  We observe that the self-energy
is transverse and independent of the temporal vector $n_\mu$.  Thus, the
renormalization structure in this method is the same as that in 
the temporal gauge calculated with some nonprincipal-value prescription.

In this work we have studied the renormalization structure of Yang-Mills theory
by the split dimensional regularization.  In this method, the dimensionality 
of the space component is $2\omega(\omega\rightarrow{3\over 2})$ and that of 
the time component is $2\sigma(\sigma\rightarrow{1\over2})$.  By using split 
dimensional regularization, we have shown that the results of integrals are 
the same as those with some nonprincipal-value prescriptions, but this 
method is seen to be considerably more straight-forward.
 
\vskip 1cm

\centerline{\bf\medium Acknowledgments}

\noindent
The authors would like to thank Prof. S.-L. Nyeo for useful discussions.
This work is supported by the National Science Council of the Republic
of China under contract number NSC 86-2112-M006-002 and NSC 86-2112-M006-005.

\ref{G. Leibbrandt, Rev. Mod. Phys. {\bf 59} (1987) 1067; {\it 
     Non-covariant Gauges}, World Scientific, Singapore, 1994.}
\ref{A. Bassetto, G. Nardelli and R. Soldati,
     {\it Yang-Mills Theories in Algebraic Non-Covariant Gauges}, World
     Scientific, Singapore, 1991.}
\ref{G. Leibbrandt and S.-L. Nyeo, Phys. Rev. {\bf D39} (1989) 1752.}
\ref{K.-C. Lee and S.-L. Nyeo, J. Math. Phys. {\bf 35} (1994) 2210.}
\ref{G. Leibbrandt and J. Williams, {\bf Guelph Math. Series 1995-151}.}

\refout
\endpage

\end